\documentclass[twocolumn,aps,floatfix,showpacs,showkeys]{revtex4}
\usepackage{color,graphicx}
\usepackage{dcolumn}
\usepackage{bm}
\usepackage{epsfig}
\usepackage{supertabular}
\usepackage[bookmarksopen]{hyperref}
\def\be {\begin{equation}}
\def\ee {\end{equation}}
\def\nn {\nonumber}
\def\bea {\begin{eqnarray}}
\def\eea {\end{eqnarray}}

\begin{document}

\title{Effect of running coupling on photons from jet - plasma 
interaction in relativistic heavy ion collisions}
\bigskip
\bigskip
\author{ Lusaka Bhattacharya}
\email{lusaka.bhattacharya@saha.ac.in}
\author{ Pradip Roy}
\email{pradipk.roy@saha.ac.in}
\affiliation{Saha Institute of Nuclear Physics, 
Kolkata - 700064, India}

\begin{abstract}
\leftskip1.0cm
\rightskip1.0cm

We discuss the role of collisional energy loss on high $p_T$ photon 
data measured by PHENIX collaboration by calculating photon yield in 
jet-plasma interaction. The phase space distribution of the 
participating jet is dynamically evolved by solving Fokker-Planck 
equation. We treat the strong coupling constant ($\alpha_s$) as function 
of momentum and temperature while calculating the drag and diffusion 
coefficients. It is observed that the quenching factor is 
substantially modified as compared to the case when $\alpha_s$ is 
taken as constant. It is shown that the data is reasonably well 
reproduced when contributions from all the relevant sources are 
taken into account. Predictions at higher beam energies relevant 
for LHC experiment have been made.

\end{abstract}

\maketitle

\section{Introduction}

Heavy ion collisions have received significant attention in recent 
years. Various possible probes have been studied in order to detect 
the signatures of quark gluon plasma (QGP). Study of direct photon 
and dilepton spectra emanating from hot and dense matter formed in 
ultra-relativistic heavy ion collisions is a field of considerable 
current interest. Electromagnetic probes have been proposed to be one 
of the most promising tools to characterize the initial state of 
the collisions~\cite{jpr}. Because of the very nature of their 
interactions with the constituents of the system they tend to 
leave the system almost unscattered. In fact, photons (dilepton as well) 
can be used to determine the initial temperature, or equivalently the 
equilibration time. These are related to the final multiplicity of 
the produced hadrons in relativistic heavy ion collisions (HIC). 
By comparing the initial temperature with the transition 
temperature from lattice QCD, one can infer whether QGP is formed 
or not.

Photons are produced at various stages of the evolution process. The 
initial hard scatterings (Compton and annihilation) of partons lead to 
photon production which we call hard photons. If quark gluon plasma 
(QGP) is produced initially, there are  QGP-photons from thermal Compton 
plus annihilation processes. Photons are also produced from different 
hadronic reactions from hadronic matter either formed initially 
(no QGP scenario) or realized as a result of a phase transition 
(assumed to be first order in the present work) from QGP. In addition 
to that there is a large background of photons coming from 
$\pi^0$ and $\eta^0$ decays. If this decay contribution is 
subtracted from the total photon yield what is left is the direct 
(excess) photons. 

These apart, there exits another class of photon emission process via 
the jet conversion mechanism 
(jet-plasma interaction)~\cite{dks} which occurs when a high energy 
jet interacts with the medium constituents via annihilation and 
Compton processes. It might be noted that this phenomenon 
(for Compton process) has been illustrated quite some time 
ago~\cite{pkrnpa} in the context of estimating photons from equilibrating 
plasma. There, it is assumed that because of the larger cross-section, 
gluons equilibrate faster providing 
a heat bath to the incoming quark-jet. A comparison of the non-equilibrium
photons (equivalent to photons from jet-plasma interaction) with the 
direct photons (thermal) shows that this contribution remains dominant 
for photons with $p_T$ upto 6 GeV. However, while evaluating jet-photon the
authors in Ref.~\cite{dks} assumes that the largest contribution to 
photons corresponds to $p_{\gamma} \sim p_q (p_{\bar q})$. This implies 
that the annihilating quark (anti-quark) directly converts into a 
photon. 
In the present work, we calculate photons from jet-plasma
interaction relaxing the above assumption and including the
energy loss of the participating jet where the strong coupling
constant ($\alpha_s$) is taken as 'running' as given in 
Ref.~\cite{prd75054031}.

The phenomena of jet quenching vis-a-vis energy 
loss have been studied by several authors \cite{jetquen} taking into 
account both collisional and radiative losses. However, in the present 
work, our main concern is to examine the role of collisional loss 
(where $\alpha_s$ is running) in the context of photon production from 
jet-plasma interaction for the following reason. 
The suppression of single electron data~\cite{phenixdil} is more than
expected which led to the re-thinking of the importance of
collisional energy loss in the context of RHIC data. A substantial
amount of work has been done to look into this issue in recent 
times~\cite{abhee05,roy06,adil,jhep04,prd75054031,prc77044904}.
Few comments about the recent developments of collisional energy loss are
in order here. 
It is argued in Ref.~\cite{nculth0302077} that the collisional energy loss
is approximately of the same order as the radiative loss.
It is also shown by Braun et. al~\cite{prd75054031} that the 
collisional energy loss increases substantially if the strong coupling is
treated as function of temperature and momentum and if, in addition to
$t$-channel process, the inverse Compton reaction is considered. 
In a most recent calculation, using a 
reduced screening mass and running coupling the collisional energy 
loss is six times larger than that with the constant 
coupling~\cite{gossiaux}. 
It explains single electron $R_{AA}$ quite well. However, it fails 
to account for the elliptic flow, $v_2$ of the electron. 
Effective resonance with LO-pQCD model~\cite{rapp} also improves the 
collisional energy loss and the single electron data is well reproduced. 
It is also important to note that only the radiative energy loss fails 
to account for the single electron data at RHIC~\cite{wicks}. On the 
other hand, the authors of Ref.~\cite{prl100072301} claims that the 
collisional energy loss is sub-leading. 
However, in order to see the effects of energy loss on jet-photon one 
should also incorporate the radiative energy loss for completeness and this 
has to be done in the same formalism in a realistic scenario.

Thus, it is apparent that the issue of the relative importance of 
the mechanism of energy loss in the context of RHIC data is not 
settled yet. We shall re-visit the importance of collisional
energy loss in the context of photons from jet-plasma interactions.
Moreover, in high temperature ($T$) effective field theory the coupling
constant, $\alpha_s$, is taken to be a function of temperature
only, which may be justified when $T >> \Lambda_{\rm {QCD}}$. However, 
in the case of relativistic heavy ion collisions, temperature is not 
the only scale, there is the momentum scale ($k$) also. One has 
to take into account the case when $k \sim T$ and treat $\alpha_s$ 
to be function of both $k$ and $T$~\cite{prd75054031}. By 
incorporating this fact it is shown that the energy loss is by 
a factor of $2-4$ more than the case when $\alpha_s$ is constant. It 
is for this purpose we concentrate on the collisional energy 
loss~\cite{abhee05,thomaplb} with the formalism given in 
Ref.~\cite{prd75054031} to calculate photons from jet-plasma interaction. 
For completeness, we also include the radiative loss in an effective way.

The organization of the paper is as follows. We give a brief 
description of jet-photon production in QGP in section IIA. The evolution of 
jet quark and photon $p_T$ distributions are discussed in sections 
IIB and IIC respectively. 
Section III is devoted to the discussions of results and 
finally, we summarize in section IV.

\section{Formalism}

\subsection{Jet-Photon Rate}

The lowest order processes for photon emission from QGP are the 
Compton scattering 
($q ({\bar q})\,g\,\rightarrow\,q ({\bar q})\,\gamma$) and 
annihilation ($q\,{\bar q}\,\rightarrow\,g\,\gamma$) process. The total 
cross-section diverges in the limit $t$ or $u \to 0$. These singularities 
have to be shielded by thermal effects in order to obtain infrared safe 
calculations. It has been argued in Ref.~\cite{kajruus} that the 
intermediate quark acquires a thermal mass in the medium, whereas the 
hard thermal loop (HTL) approach of Ref.~\cite{Brapi} shows that very 
soft modes are suppressed in a medium providing a natural cut-off 
$k_c \sim gT$. We assume that the singularities can be shielded by 
the introduction of thermal masses for the participating partons. 
Apart from the thermal interactions of the plasma partons, interaction 
of a leading jet parton with the plasma was found to be a very 
important source of photons.

The differential photon production rate for this process is given by:
\begin{eqnarray} 
&&E\frac{dR}{d^3p}=\frac{{\mathcal{N}}}{2(2\pi)^3} 
\int \frac{d^3p_1}{2E_1(2\pi)^3}\frac{d^3p_2}{2E_2(2\pi)^3}
\frac{d^3p_3}{2E_3(2\pi)^3}
f_{jet}({\bf{p_1}})\nonumber\\
&&\times f_2(E_2)
(2\pi)^4\delta(p_1+p_2-p_3-p)
|{\mathcal{M}}|^2 (1\pm f_3({E_3}))
\label{photonrate}
\end{eqnarray}
where, $|{\mathcal{M}}|^2$ represents the spin averaged matrix element 
squared for one of those processes which contributes in the photon rate 
and ${{\mathcal N}}$ is the degeneracy factor of the corresponding 
process. $f_{jet}$, $f_2$ and $f_3$ are the initial state and final 
state partons. $f_2$ and $f_3$ are the Bose-Einstein or Fermi-Dirac
distribution functions.
\begin{eqnarray}
f_2(E_{2,3}) = \frac{1}{exp(E_{2,3}/kT)\pm 1}
\end{eqnarray}

\subsection{Fokker - Planck Equation: Parton transverse 
momentum spectra}

In the photon production rate (from jet-plasma interaction) one of the 
collision partners is assumed to be in equilibrium and the other 
(the jet) is executing random motion in the heat bath provided by quarks 
(anti-quarks) and gluons. Furthermore, the interaction of the jet is 
dominated by small angle scattering. In such scenario the evolution 
of the jet phase space distribution is governed by Fokker-Planck 
(FP) equation where the collision integral is approximated by 
appropriately defined drag and diffusion coefficients.

As mentioned already in the introduction that the quark jet here is 
not in equilibrium. Therefore the corresponding distribution function 
($f_{jet}$) that appears in Eq.~(\ref{photonrate}) is calculated by 
solving the FP equation. The energy loss is represented by the drag 
coefficient (see later). The FP equation, can be derived from Boltzmann 
equation if one of the partners of the binary collisions is in thermal 
equilibrium and the collisions are dominated by the small angle 
scattering involving soft momentum 
exchange~\cite{roy06,alamprl94,svetitsky,moore05,ducati,rajuprc01,rapp}. 
For a longitudinally expanding plasma, FP equation reads~\cite{roy06,baym}:
\begin{eqnarray}
\left (\frac{\partial}{\partial t}
-\frac{p_z}{t}\frac{\partial}{\partial p_z}\right )f({\bf p_T},p_z,t)
=\frac{\partial}{\partial p_i}A_i({\bf p}) f({\bf p}) +\nn\\ 
\frac{1}{2}
\frac{\partial}{\partial p_i \partial p_j}[B_{ij}({\bf p})f({\bf p})], 
\label{fpexp}
\end{eqnarray}
where~\cite{roy06}
\begin{eqnarray}
A_i&=&\frac{\nu}{16p(2\pi)^5}\int 
\frac{d^3{k^\prime}}{{k^\prime}}
\frac{d^3k}{k}
\frac{d^3{q}}{{p^\prime}}d\omega q_i
|{\cal{M}}|_{t\rightarrow 0}^2 f(k) (1+f(k{^\prime}))\nn\\ 
&&\delta^3({\bf q}-{\bf k^\prime}+ {\bf k})
\delta (\omega-{\bf v_{k^\prime}\cdot q})
\delta (\omega-{\bf v_k\cdot q})
\label{eq:drag1}
\end{eqnarray}
\begin{eqnarray}
B_{ij} =\frac{\nu}{16p(2\pi)^5}\int 
\frac{d^3{k^\prime}}{{k^\prime}}
\frac{d^3k}{k}
\frac{d^3{q}}{{p^\prime}}d\omega q_iq_j
|{\cal{M}}|_{t\rightarrow 0}^2 f(k)\nn\\ \times (1+f(k{^\prime})) 
\delta^3({\bf q}-{\bf k^\prime}+ {\bf k})
\delta (\omega-{\bf v_{k^\prime}\cdot q})
\delta (\omega-{\bf v_k\cdot q}),
\label{eq:diffusion}
\end{eqnarray}
Here $\nu$ is the appropriate degeneracy factor. 
Note that the coefficient $A_i$ is related the drag coefficient 
$\eta$ by $A_i = \eta p_i$, where $\eta =(1/E)dE/dx$.

Now, $B_{ij}$ can be decomposed into  longitudinal and transverse 
components:
\bea
B_{ij}=B_t (\delta_{ij}-\frac{p_i p_j}{p^2}) + B_l \frac{p_ip_j}{p^2}
\eea
Explicit calculation shows that the off diagonal components of 
$B_{ij}$ vanish and we have,
\bea
B_{t,l}&=&\frac{\nu}{(2\pi)^4}
\int\frac{d^3kd^3qd\omega}{2k2k^\prime 2p 2p^\prime} 
\delta(\omega-{\bf v_p\cdot q})
\delta(\omega-{\bf v_k\cdot q})\nn\\
&&\langle {\cal M} \rangle_{t\rightarrow 0}^2
f(k)[1+f(k)+\omega\frac{\partial f}{\partial k}]
q_{t,l}^2,
\label{eq:blt}
\eea
where, 
$B_l=d\langle(\Delta p_z)^2\rangle/dt$,
$B_t=d\langle(\Delta p_T)^2\rangle/dt$, represent diffusion constants 
along parallel and perpendicular directions of the propagating partons. 
Evidently, $A_i$ ($B_{t,l}$) is infrared singular. Such divergences do 
not arise if close and distant collisions are treated separately. For 
very low momentum transfer the concept of individual collision breaks 
down and one has to take collective excitations of the plasma into account. 
Hence there should be a lower momentum cut off above which 
bare interactions might be considered. While for soft collisions 
medium modified hard thermal loop corrected propagator should be 
used~\cite{thoma91,abhee05}. It is evident that Eq.~(\ref{eq:drag1}) 
actually gives $dE/dt$ or the energy loss rate~\cite{abhee05} that can 
be related to the drag coefficient.
%
%
\vspace{0.5cm}
\begin{figure}[htb]
\epsfig{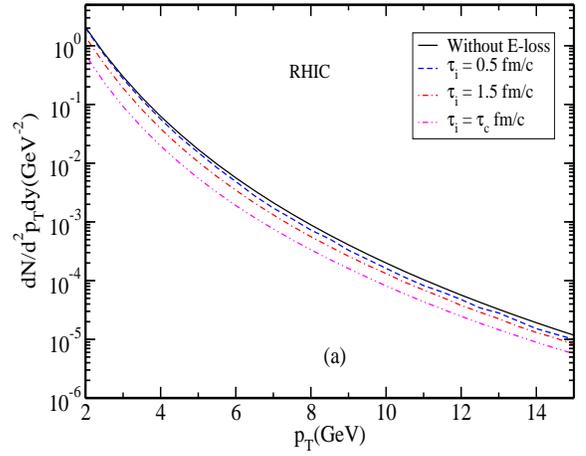}
\vskip 0.5in
\epsfig{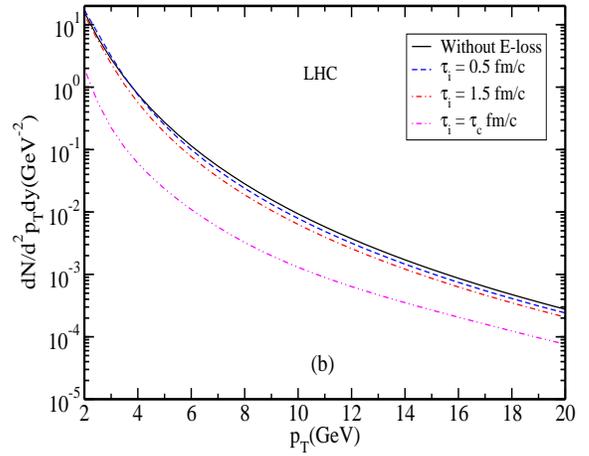}
\caption{(Color online) $p_T$ distribution of parton at (a) RHIC energy 
(${T_i}= 0.446$ GeV and ${\tau_i}= 0.147$ fm/c) and at 
(b) LHC energy ($T_i=0.897$ GeV, $\tau_i=0.073$ fm/c).}
\label{fig_ptdist}
\end{figure}

However, in the above treatment, the infra-red cut-off is fixed by 
plasma effects, where only the medium part is considered, completely 
neglecting the vacuum contribution leading to ambiguity in the energy 
loss calculation. If the latter part is taken into account the strong 
coupling should be running. Thus for any consistent calculation one has 
to take into consideration this fact. In that case 
$\alpha_s=\alpha_{s}(k,T)$ ($k=\sqrt{|\omega^2-q^2|}$ in this case), 
and the above integrals must be evaluated numerically where the infra-red 
cut-off is fixed by Debye mass to be solved self-consistently:
\be
{m_{D}}(T) = 4\pi\,\left(1+\frac{N_F}{6}\right)\,\alpha_s(m_D(T),T)T^2
\label{eq:dmass}
\ee
We reiterate that the matrix elements in Eqs. (\ref{eq:drag1}) 
and (\ref{eq:blt}) contains the strong coupling which we take as 
running, i. e. $\alpha_s= \alpha_s (\sqrt{|\omega^2-q^2|},T)$. 
We chose the following parameterization of $\alpha_s$ which respects 
the perturbative ultra-violet (UV) behavior and the 3D infra-red 
(IR) point~\cite{prd75054031}:
\begin{eqnarray}
&&\alpha_s (k,T)= \frac{u_1 \frac{k}{T}}{1+exp(u_2\frac{k}{T}-u_3)}\nn\\
&+& \frac{v_1}{(1+exp(v_2 \frac{k}{T}-v_3))(ln(e+ (\frac{k}{\lambda_s})^a
+(\frac{k}{\lambda_s})^b)},
\end{eqnarray}
with $k = \sqrt{|\omega^2-q^2|}$ in this case. 
The parameters $a$, $b$ and $\lambda_s$ are given by
$a=9.07$, $b=5.90$ and $\lambda_s=0.263$ GeV. 
For the limiting behavior ($k << T$) of the coupling we choose,
\bea
u_1={\alpha^*}_{3d}(1 + exp(-u_3))
\eea 
Here ${\alpha^*}_{3d}$ and ${\alpha^*}_s$ denote the values of the IR 
fixed point of $SU(3)$ Yang-Mills theory in $d=3$ and $d=4$ dimensions, 
respectively. The remaining four parameters ($u_2=5.47, u_3=6.01, v_2=10.13$ 
and $v_3=9.27$) fit the numerical results for pure Yang-Mills theory 
obtained from the RG equations in Ref.~\cite{Braun_Gies}.

So far we have discussed about the collisional energy loss which, although,
dominates at lower energies, is not the only mechanism 
of energy loss. As the energy increases radiative energy loss
starts to dominate and hence cannot be neglected.
In case of jet-photon production, since the photon energy is
almost equal to the jet energy, one has to include the radiative
loss to account for the high $p_T$ photons. However, in order to
see the effects of both the collisional and radiative energy losses, 
one must develop a formalism in which both the mechanisms can be taken 
into account in a consistent manner.
The two mechanisms are not entirely independent, i.e., the collisional
loss may influence the radiative loss. Thus both should be included to
calculate transport coefficients. Since there is no rigorous way
to implement this, the approximate way is to define  effective
drag (diffusion) in the following manner:
\begin{eqnarray}
\eta& =& \eta_{\rm coll} + \eta_{\rm rad}\nonumber\\
&=&\frac{1}{E}\left[\left(\frac{dE}{dx}\right)_{\rm coll}+
\left(\frac{dE}{dx}\right)_{\rm rad}\right]
\end{eqnarray}
where $\left(\frac{dE}{dx}\right)_{\rm coll}$ can be calculated from
Eq. (\ref{eq:drag1}) keeping in mind that $\eta p_i=A_i$.
The collisional (differential) energy loss $(dE/dx)_{\rm coll}$ can be 
calculated from Eq.~(\ref{eq:drag1}) using the method described in 
\cite{roy06}. For running $\alpha_s$ it is given by,
\begin{eqnarray}
\left(\frac{dE}{dx}\right)_{\rm coll} = 
\frac{4}{3}\pi(1+\frac{N_f}{6})T^2 {\int^s}_{{m_D}^2}
d|t|\frac{{\alpha_s}^2(\sqrt{|t|}, T)}{|t|^2}|t| \nonumber\\
\label{coll}
\end{eqnarray}
Where, $s=2ET$, $E$ is the energy of the incident quark. 
The radiative energy loss is given by,
\begin{equation}
\left(\frac{dE}{dx}\right)_{\rm rad}=
\frac{C_F\alpha_s(E,T)}{N(E)}\frac{L\mu^2}{\lambda_g}
\ln(\frac{E}{\mu}),
\label{rad3}
\end{equation}
where $N(E)$ is an energy dependent factor and $N(E\rightarrow \infty)=4$ if
kinematic bounds are neglected~\cite{gyulassy00prl}.
It is important to point out here that
$N(E)=7.3, 10.1, 24.4$ for $E=500, 50, 5 $ GeV respectively and 
$N(E\rightarrow\infty)=4$ (see the Ref.~\cite{gyulassy00prl}). 
$L$ is the distance traversed by the jets in the plasma.
Similarly we can define the effective diffusion coefficients.

Having known the drag and diffusion, we solve the FP equation using 
Green's function techniques: If $P(\vec{p},t|\vec{p_0},t_i)$ is a 
solution to Eq. (\ref{fpexp}) with the initial condition
\begin{equation}
P(\vec{p},t=t_i|\vec{p_0},t_i) = \delta^{(3)}(\vec{p}-\vec{p_0})
\end{equation}
the full solution with an arbitrary initial condition can be
obtained as~\cite{moore05}
\begin{equation}
f(t,\vec p) = \int d^3{\vec p_0} P(\vec{p},t|\vec{p_0},t_i)f_0(\vec p_0)
\end{equation}
where for the initial condition $f(t=0, \vec p)=f_0(p_0)$ and
$P(\vec{p},t|\vec{p_0},t_i)$ is the Green's function of the partial
differential Eq. (2).

We assume here that the plasma expands only longitudinally 
(Bjorken expansion scenario~\cite{bj}). The reason is the following. 
The transverse expansion will have two effects on the parton energy 
loss: (i) The expanding geometry will increase the duration of 
propagation, (ii) the same expansion will cause the parton density 
to fall along its path. These two effects partially compensate each 
other and the energy loss is almost the same as in the case without 
transverse expansion~\cite{plb526}. 
Since we are considering the central rapidity 
region $(|\eta|<0.35)$ the arguments given above is not applicable 
in the case of longitudinal expansion scenario.

The solution with an arbitrary initial momentum distribution can now be 
written as~\cite{rapp},
\begin{equation}
E\frac{dN}{d^3p} = \int\,d^3p_{0}\,
P(\vec{p},t|\vec{p_0},t_i) E_0\frac{dN}{d^3p_0}\,
\label{rapp_prc71}
\end{equation}
We use the initial parton $p_T$ distributions (at the formation 
time $t_i$) taken from~\cite{dks,muller}:
\begin{eqnarray}
\frac{dN}{d^2p_{0T}dy_0}|_{y_0=0}=
\frac{K N_0}{(1+\frac{p_{0T}}{\beta})^\alpha},
\label{inpt}
\end{eqnarray}
where $K$ is a phenomenological factor ($\sim 1.5 - 2$) which takes 
into account the higher order effects. The values of the 
parameters are listed in Table.~\ref{parameter}.
\begin{table}[h]
\begin{center}
\begin{tabular}{|c|c|c|c|c|}
\hline\hline
& \multicolumn{2}{|c|}{RHIC} & \multicolumn{2}{|c|}{LHC} \\\cline{2-5}
& $q$ & $\bar q$ & $q$ & $\bar q$ \\\hline\hline
 $N_0~[1/GeV^2]$ & $5.0\times 10^2$& $1.3\times 10^2$& $1.4\times 10^4$&
$1.4\times 10^5$ \\ \hline
$\beta~[GeV]$ & 1.6& 1.9& 0.61& 0.32 \\ \hline
$\alpha$ & 7.9& 8.9& 5.3& 5.2 \\\hline\hline
\end{tabular}
\end{center}
\caption{Parameters for initial parton $p_T$ distribution.} 
\label{parameter}
\end{table} 
We note that the parametric form of Eq. (\ref{inpt}) may not represent 
the true picture of the jet $p_T$ distribution. In recent years, more 
sophisticated calculations have been done using different 
parameterizations of the parton distribution functions. The $p_T$ 
distribution used here might differ substantially  from these 
calculations and the results presented here is correct upto a factor 
that might come from using the more state of the art calculation.

\subsection{Space time evolution}

In order to obtain the space-time integrated rate we first note that 
the phase space distribution function for the incoming jet in the mid 
rapidity region is given by (see Ref.~\cite{prc72} for details)
\begin{eqnarray}
f_{jet}(\vec r,\vec p,t^{\prime})|_{y=0}&=&
\frac{(2\pi)^3\mathcal{P}(|{\vec w}_r|)~t_i}{\nu_q \sqrt{{t_i}^2-{z_0}^2}}
\frac{1}{p_{T}}\nonumber\\
&\times&\frac{dN}{d^2{p_{T}}dy}(p_T,t^{\prime})\delta(z_0)
\label{jetp}
\end{eqnarray}
where $\frac{dN}{d^2{p_{T}}dy}(p_T,t^{\prime})$ can be obtained
from Eq. (\ref{rapp_prc71}). $t_i$ is the jet formation time and 
$\nu_q$ is the spin-color degeneracy factor. $z_0$ is the jet formation 
position in the direction of QGP expansion and 
$\mathcal{P}(|{\vec w}_r|)$ is the initial jet production probability 
distribution at the initial radial position ${\vec w}_r$ in the 
plane $z_0=0$, where
\begin{eqnarray}
|{\vec w}_r|&=&(\vec {r}-(t^{\prime}-t_i)~
\frac{\vec{p}}{\vec{|p|}})\cdot \hat{r}
\nonumber\\
&=&\sqrt{(r\cos{\phi}-t^{\prime})^2+r^2\sin^2{\phi}}  \ \ {\rm for}\ t_i\sim 0
\end{eqnarray}
and $\phi$ is the angle in the plane $z_0=0$ between the direction of the
photon and the position where this photon has been produced. We assume 
the plasma expands only longitudinally. Thus using 
$d^4x=rdrdt^{\prime}d\phi dz$ and the expression for $f_{jet}$ 
from Eq.~\ref{inpt} we obtain the transverse momentum distribution 
of photon as follows~\cite{prc72,kapusta}:
\begin{eqnarray}
\frac{dN^{\gamma}}{d^2p_Tdy}&=&\int d^4x ~ \frac{dN^{\gamma}}{d^4xd^2p_Tdy} 
\nonumber\\
&=&\frac{(2\pi)^3}{\nu_q}{\int_{t_i}}^{t_c}dt^{\prime}
{\int_0}^R rdr \int d\phi\mathcal{P}(\vec {w_r})\nonumber\\
&\times& \frac{{\mathcal{N}_i}}{16(2\pi)^7E_{\gamma}}\int 
d{\hat s}d{\hat t} |\mathcal{M}_i|^2\int {dE_1 dE_2}\nonumber\\
&\times&\frac{1}{p_{1T}}\frac{dN}{dp_{1T}^2dy}(p_{1T},t^\prime)\frac{f_{2}(E_2)
(1\pm f_3(E_3))}{\sqrt{a{E_2}^2+2bE_2+c}}\nn\\
\label{last}
\end{eqnarray}
$f_{jet}$ is the distribution function of the jet quark 
(see Eq.~(\ref{jetp})) and rest of the distribution functions 
i.e $f_2,f_3$ are Fermi-Dirac or Bose-Einstein distributions. 
$\phi$ dependence occurs only in $\mathcal{P}(\vec {w_r})$. 
So the $\phi$ integration can be done analytically as in 
Ref.~\cite{prc72}. The temperature profile is taken 
from Ref.~\cite{prc72}.

Besides the thermal photons from QGP and hadronic matter we also
calculate photons from initial hard scattering from the reaction of the
type $h_A\,h_B\,\rightarrow\,\gamma\,X$ using perturbative QCD. We include
the transverse momentum broadening in the initial state
partons~\cite{wong,owens}. The cross-section for this process can then 
be written in terms of elementary parton-parton cross-section multiplied 
by the partonic flux which depends on the parton distribution functions 
(PDF) for which we take CTEQ parameterization~\cite{cteq}. A phenomenological 
factor $K$ is used to take into account the higher order effects. We also 
include photons from fragmentation process.

\section{Results}

%
\begin{figure}[htb]
\centerline{\epsfig{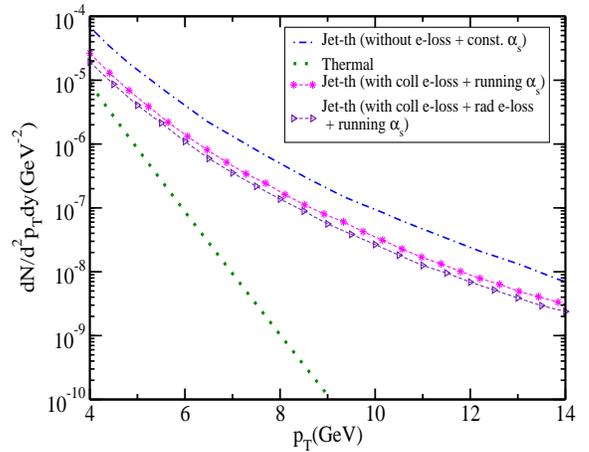}}
\caption{(Color online) $p_T$ distribution of photons at RHIC energy 
with ${T_i}= 0.446$ GeV and ${\tau_i}= 0.147$ fm/c. 
The violet (magenta) curve denotes the photon yield from 
jet-plasma interaction with collisional (collisional + radiative) 
energy loss. The blue curve corresponds to the case without any energy 
loss and the green curve represents the thermal contribution.}
\label{compare}
\end{figure}
%

To obtain the quark momentum distribution we use 
Eqs. (\ref{fpexp}), (\ref{eq:drag1}), (\ref{eq:blt}), (\ref{eq:dmass}) 
and (\ref{rapp_prc71}). The transverse momentum distributions of quarks 
are shown in Fig.~\ref{fig_ptdist} for different times (proper) at 
RHIC and LHC energies respectively where the initial distributions 
are taken from Eq.~(\ref{inpt}). It is observed that the spectra are 
more reduced as the time increases. It is generally assumed that quarks 
fragment into hadrons around $\tau_c$ 
(the begining of the hadronic phase) where the quenching factor is 
the largest. At LHC energies (see Fig.~\ref{fig_ptdist}b) this factor 
is more as the temperature in this case is large compared to RHIC energies. 
It is seen from Eq.~(\ref{last}) the photon $p_T$ distribution is directly 
proportional to the quark $p_T$ spectra. Thus the photon yield will 
be affected as we shall see in the following.
%
%
\begin{figure}[htb]
\bigskip
\centerline{\epsfig{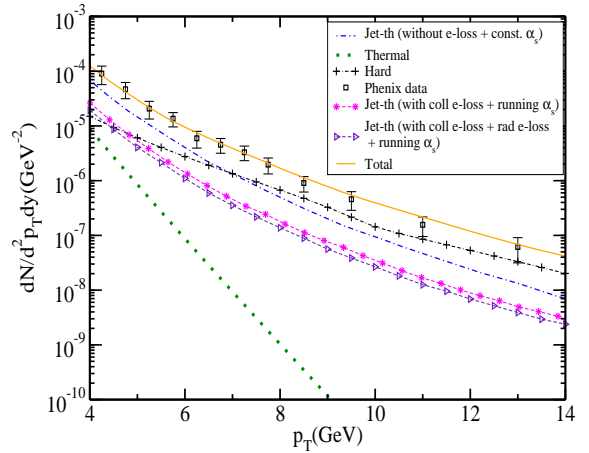}}
\caption{(Color online) $p_T$ distribution of photons at RHIC energy 
with ${T_i}= 0.446$ GeV and ${\tau_i}= 0.147$ fm/c. 
The magenta (blue) curve denotes the photon yield from jet-plasma interaction 
running (constant) $\alpha_s $.  The black (green) curve corresponds 
to hard (thermal) photons. The orange represents the total photon yield 
compared with the Phenix measurements of photon data~\protect\cite{phenix}.}
\label{fig_rhic446}
\end{figure}
%

In order to obtain the photon $p_T$ distribution we numerically integrate 
Eq. (\ref{last}) using Eq. (\ref{rapp_prc71}). 
The results for jet-photons for RHIC energies are plotted in 
Fig.~\ref{compare} where we have taken $T_i = 446$ MeV and 
$t_i = 0.147$ fm/c. As indicated earlier, the radiative energy loss
starts dominating at higher energies of the jet, we include this in the
calculation of photon $p_T$ distribution. We find that the yield
is decreased with the inclusion of both the energy loss
mechanisms as compared to the case when only collisional energy loss
is considered. It is to be noted that when one considers
collisional energy loss alone the yield with constant
$\alpha_s$ is more compared to the situation when running
$\alpha_s$ is taken into account. This is due to the fact that
the energy loss in the later case is more~\cite{prd75054031}. 
On the other hand, with the inclusion of the radiative loss 
the yield decreases further.

%
%
\vskip 0.7cm
\begin{figure}[htb]
\bigskip
\centerline{\epsfig{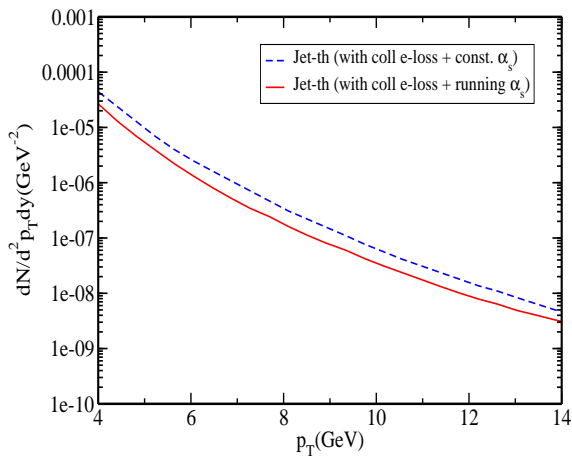}}
\caption{(Color online) $p_T$ distribution of jet-th photons at RHIC energy 
with ${T_i}= 0.446$ GeV and ${\tau_i}= 0.147$ fm/c. 
The red (blue) curve denotes the photon yield coming from jet-plasma 
interaction with collisional energy loss + running (constant) $\alpha_s $.}  
\label{fig_rhic446_coll-eloss}
\end{figure}
%

In order to compare our results with high $p_T$ photon data measured by 
the PHENIX collaboration~\cite{phenix}, we have to evaluate the 
contributions to the photons from other sources, that might contribute 
in this $p_T$ range. In Fig.~\ref{fig_rhic446} the results for jet-photons 
corresponding to the RHIC energies are shown, where we have taken 
$T_i = 446$ MeV and $t_i = 0.147$ fm/c. The individual contributions 
from hard and bremsstrahlung processes~\cite{Owens} are also shown for 
comparison. 
These are estimated using the formalism given in Ref. \cite{Owens}. 
The total yield comprises of photons from jet-plasma interaction 
(with energy loss), hard and bremsstrahlung processes, thermal Compton and
annihilation processes.  
We also show in a separate plot (in Fig.~\ref{fig_rhic446_coll-eloss}) 
photons from jet-plasma interaction corresponding to the cases with constant 
$\alpha_s$ and running $\alpha_s$ with collisional energy loss alone. 
It is observed that the spectra in the case of collisional energy loss 
with running coupling is depleted by a factor {$1.7 - 2$} compared to the 
case where the strong coupling is constant. This is expected as the 
energy loss is more in the former case. The yield 
further reduces when both the mechanisms of energy loss are included. 
The total photon yield consisting of jet-photon, photons from initial 
hard collisions, jet-fragmentation and thermal photons is compared with 
the PHENIX photon data~\cite{phenix}. It is seen that the data is well 
reproduced in our model (see Fig.~\ref{fig_rhic446}). 
%
%
\begin{figure}[htb]
\bigskip
\bigskip
\centerline{\epsfig{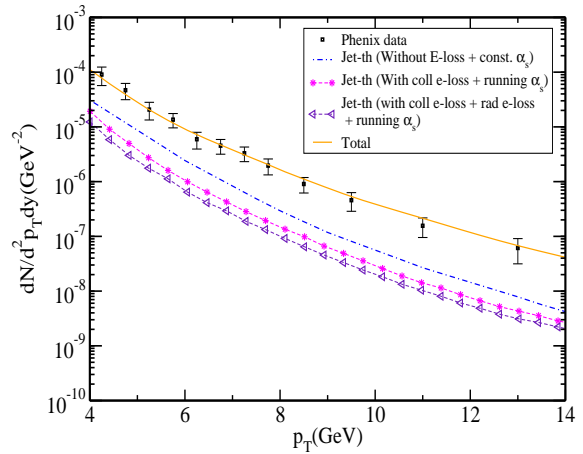}}
\caption{(Color online) $p_T$ distribution of photons at RHIC energies with 
$T_i= 0.350$ GeV and $\tau_i= 0.25$ fm/c. The orange line corresponds
to total photon yield from all the sources as in 
Fig.~\protect\ref{fig_rhic446} with both energy losses 
(collisional + radiative) included in the jet-photon contribution. }
\label{fig_rhic350}
\end{figure}
%
%
%
To cover the uncertainties in the initial conditions for a given 
beam energy, we consider another set of initial conditions at a lower 
temperature $T_i=0.350$ GeV and somewhat later initial time of 
$\tau_i=0.25$ fm/c. The yield for this set is shown 
in Fig.~\ref{fig_rhic350}. We see that the data is reproduced 
reasonably well. 

%
%
\begin{figure}[htb]
\bigskip
\centerline{\epsfig{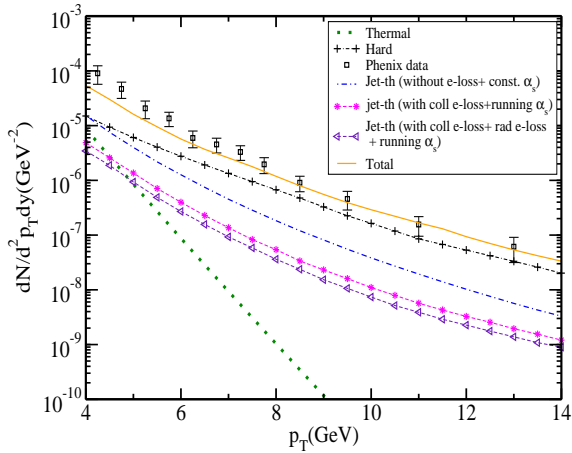}}
\caption{(Color online) $p_T$ distribution of photons at RHIC energy 
with ${T_i}= 0.236$ GeV and ${\tau_i}= 0.5$ fm/c. The magenta (blue) 
curve denotes the photon yield from jet-plasma interaction running 
(constant) $\alpha_s $.  The black (green) curve corresponds to hard 
(thermal) photons. The orange represents the total yield compared with 
the Phenix measurements of photon data~\protect\cite{phenix}.}
\label{fig_rhic446_dndy600}
\end{figure}
%
In Fig.~\ref{fig_rhic446_dndy600} we plot the $p_T$ distribution of 
photons for the RHIC energy ($T_i=0.236$ GeV and $\tau_i=0.5$ fm/c) 
for a lower value of $dN/dy=600$. It is clearly visible from 
Fig.~\ref{fig_rhic446_dndy600} that we can not explain Phenix photon 
data satisfactorily in the $p_T$ range $4 - 8$ GeV. For the higher 
$p_T$ range Phenix photon data is well reproduced.

\begin{figure}[htb]
\bigskip
\bigskip
\centerline{\epsfig{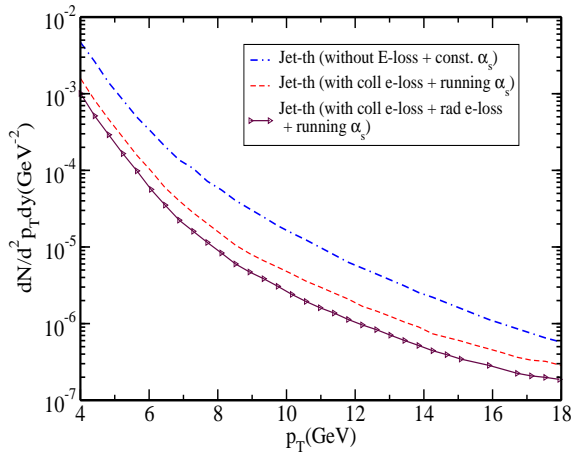}}
\caption{(Color online) Same as Fig.(\protect\ref{fig_rhic446}) at LHC 
energy ($T_i= 0.897$ GeV and $\tau_i= 0.073$ fm/c).}
\label{fig_lhc}
\end{figure}

We also consider the high $p_T$ photon production at LHC energies. 
The contributions from various sources are shown  in Figs.~\ref{fig_lhc} 
where the jet-plasma contribution is calculated with running coupling 
constant ($\alpha_s$) (considering both the collisional and 
radiative energy losses).
Since the initial temperature in this case is higher, the plasma lives 
for longer time. Thus the energy loss suffered by the parton is more. 
As a result, the difference between the cases with and without energy 
loss is slightly more than what is obtained at RHIC. It is observed 
that due to the inclusion of radiative energy loss along with 
collisional energy loss the jet photon yield is suppressed significantly. 
Also due to the inclusion of running coupling constant the jet photon yield 
is suppressed by a factor of $3 - 4$. 
%

\section{Summary}

We have calculated the transverse momentum distribution of
photons from jet plasma interaction with running coupling, i. e. 
with $\alpha_s = \alpha_s(k,T)$ where we have included both collisional 
and radiative energy losses. It is found that the assumption made in 
Ref.~{\cite{dks}} while calculating photons from jet-plasma interactions 
may not be good at LHC energies as we observe a difference is by a factor 
of $2 - 3$. Using running coupling we find that the depletion in the 
photon $p_T$ spectra is by a factor of $2 - 2.5$ more as compared to 
the case with constant coupling for RHIC energies. This is due to the 
fact that the energy loss (and hence drag and diffusion coefficients) 
is more by similar factor in the case of running coupling. 
Phenix photon data have been contrasted with the present calculation 
and the data seem to have been reproduced well in the low 
$p_T$ domain. The energy of the jet quark to produce photons in this 
range ($4 < p_T < 14$) is such that collisional energy loss plays 
important role here. It is shown that inclusion of radiative energy 
loss also describes the data reasonable well.
 
To check the sensitivity with the initial conditions, we consider two 
sets of initial conditions. In both the cases the data can be described 
quite well. This is due to the fact that both the initial conditions
corresponds to the same $dN/dy = 1150$.

As we validate our model through the description of Phenix photon 
data we also predict the high $p_T$ photon yield that might be expected 
in the future experiment at LHC. We notice that the inclusion 
of the radiative energy loss further reduces the yield at high $p_T$. 
It is observed that the contribution from jet-plasma interaction is 
slightly more reduced as compared to the RHIC case as the initial 
temperature is higher at LHC. 

We do not consider transverse expansion as the energy loss of the partons
remains just about the same as the case without transverse expansion.
Finally, we conclude by noting that the role of running coupling 
constant should be explored in the context of other observables 
such as thermal photons, dileptons and so on.

\end{document}